\documentclass[journal]{IEEEtran}

\usepackage{cite}
\usepackage{amsmath}
\usepackage{graphicx}

\newcommand{\ud}{\mathrm{d}}
\newcommand{\eqn}[1]{(\ref{#1})}
\newcommand{\fig}[1]{Fig. \ref{#1}}
\newcommand{\Fig}[1]{Figure \ref{#1}}

\begin{document}

\title{Comparison of Nonlinear Phase Noise and Intrachannel Four-Wave-Mixing for RZ-DPSK Signals in Dispersive Transmission Systems}

\author{Keang-Po Ho,~\IEEEmembership{Senior Member,~IEEE} and Hsi-Cheng Wang%
\thanks{Manuscript received January 8, 2005, revised ??, 2005. %
This research was supported in part by the National Science Council of Taiwan under Grant NSC-93-2213-E-002-061, NSC-93-2219-E-002-007, and NSC-93-2219-E-002-008}
\thanks{K.-P. Ho is with the Institute of Communication Engineering and Department of Electrical Engineering, National Taiwan University, Taipei 106, Taiwan.(Tel: +886-2-3366-3605, Fax: +886-2-2368-3824, E-mail: kpho@cc.ee.ntu.edu.tw)}
\thanks{H.-C. Wang is  with the Institute of Communication Engineering, National Taiwan University, Taipei 106, Taiwan.}
}

\markboth{IEEE Photonics Technology Letters}{K.-P. Ho: Comparison of Nonlinear Phase Noise and Intrachannel Four-Wave-Mixing for RZ-DPSK Signals in Dispersive Transmission Systems}

\maketitle

\begin{abstract}
Self-phase modulation induced nonlinear phase noise is reduced with the increase of fiber dispersion but intrachannel four-wave-mixing (IFWM) is increased with dispersion.
Both degrading DPSK signals, the standard deviation of nonlinear phase noise induced differential phase is about three times that from IFWM even in highly dispersive transmission systems. 
\end{abstract}


\section{Introduction}

\PARstart{R}{ecently}, differential phase-shift keying (DPSK) signal has been studied extensively for long-haul lightwave transmissions \cite{gnauck02, rasmussen03, cai04, charlet04a}.
Mostly for 40-Gb/s systems, DPSK signal has 3-dB receiver sensitivity improvement and provides good tolerance to fiber nonlinearities than on-off keying.
Most DPSK experiments use return-to-zero (RZ) pulse and launch a pulse train with phase modulated to each RZ pulse.

The interaction of fiber Kerr effect with amplifier noise induces nonlinear phase noise \cite{gordon90, kim03, ho0405, ho03sta}, or more precisely, self-phase modulation (SPM) induced nonlinear phase noise.
Added directly to the signal phase, as shown later, nonlinear phase noise is the major degradation for DPSK signals.
 
When RZ pulse broadens by chromatic dispersion and overlaps with each other, the pulse-to-pulse collision gives intrachannel cross-phase modulation (IXPM) and four-wave-mixing (IFWM) \cite{shake98, essiambre99}.
While IXPM has no effect on DPSK signals, IFWM adds ghost pulses to each DPSK RZ pulse \cite{mecozzi00, mecozzi01a, abo2, wei03fwm, ho0504}. 

For RZ-DPSK signals, the variance of the nonlinear phase noise is derived here analytically, to our knowledge, the first time.
Comparing with the IFWM variance from \cite{wei03fwm, ho0504}, the phase noise standard deviation (STD) from nonlinear phase noise is about three times larger than that from IFWM even at highly dispersive transmission systems.

\section{Nonlinear Phase Noise for RZ Pulses}

For a comparison to IFWM, nonlinear phase noise is evaluated based on the model of \cite{mecozzi00, mecozzi01a, abo2, wei03fwm, ho0504}.
Assumed a Gaussian pulse with an initial $1/e$ pulse width of $T_0$, the $k$th pulse along the fiber is 
\begin{equation}
u_k(z, t) = \frac{A_k T_0 e^{-\alpha z/2}} {(T_0^2 - j \beta_2 z)^{1/2}} 
        \exp\left[ - \frac{(t-kT)^2}{2 (T_0^2 - j \beta_2 z)}\right],
\label{ukzt}
\end{equation}
\noindent where $A_k = \pm A_0$ is the pulse amplitude modulated by either $0$ or $\pi$ phases, $\beta_2$ is the coefficient of group velocity dispersion, $T$ is the bit interval, and $\alpha$ is the fiber attenuation coefficient.
Due to fiber Kerr effect, from the model of \cite{mecozzi00, abo2, mecozzi01a}, there is a nonlinear force of $j \gamma u_k u_l u^*_m$ from the collision of the $k$-, $l$-, and $m$-th pulses, where $\gamma$ is the fiber nonlinear coefficient.
The overall ghost pulse is equal to 
\begin{equation}
j \gamma \int_{0}^L \left[u_k(z, t) u_l(z, t) u^*_m(z, t)\right] \otimes h_{-z}(t) \ud z,
\label{ifwmmodel}
\end{equation} 
\noindent where $\otimes$ denotes convolution, and $L$ is the fiber length.
The impulse response of $h_{-z}(t)$ provides dispersion compensation for $h_z(t)$ where $h_z(t)$ is the impulse response for fiber chromatic dispersion, the corresponding frequency response is $H_z(\omega) = \exp( j \beta_2 z \omega^2/2)$.

To be consistent with the model for IFWM of \eqn{ifwmmodel}, for the pulse of $u_0(z, t)$, the SPM-induced nonlinear force including amplifier noise of $n(z, t)$ is equal to 
\begin{equation}
j \gamma [u_0(z, t) + n(z, t)] \left| u_0(z, t) + n(z, t)\right|^2.
\end{equation}
\noindent For the signal, nonlinear force is $j \gamma u_0 |u_0|^2$ or that of \eqn{ifwmmodel} with $k = l = m = 0$.
The nonlinear force associated with nonlinear phase noise has two different terms of 
\begin{equation}
2 j \gamma |u_0(z, t)|^2 n(z, t), \mbox{~~~and~~~} j \gamma u^2_0(z, t) n^*(z, t),
\end{equation}
\noindent when all quadratic or higher-order terms of the noise are ignored.
For $2 j \gamma |u_0(z, t)|^2 n(z, t)$, the nonlinear force corresponding to \eqn{ifwmmodel} is equal to
\begin{equation}
\Delta u_n(t) = 2 j \gamma \int_{0}^L \left[|u_0(z, t)|^2 n(z, t)\right] \otimes h_{-z}(t) \ud z.
\end{equation}

\begin{figure}
\centerline{
\includegraphics[width = 0.40 \textwidth]{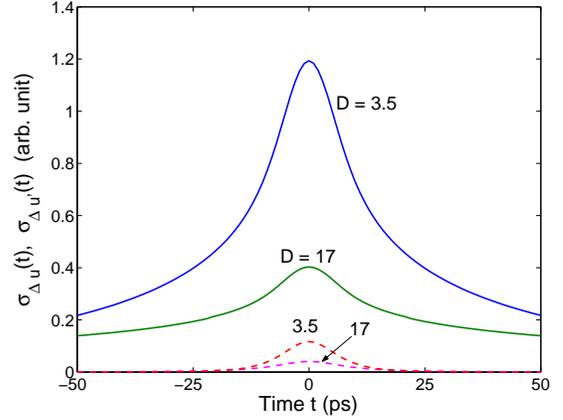}}
\caption{The temporal distribution of nonlinear force due to the beating of signal with noise.
The solid lines are $\sigma_{\Delta u_n}(t)$ and the dashed-lines are $\sigma_{\Delta u_n^\prime}(t)$.
}
\label{figsigmaut}
\end{figure}

At the input of the fiber, we assume that $E\left\{n(0, t + \tau) n^{*}(0, t) \right\} = 2\sigma_n^2 \delta(\tau)$ as a white noise, where $\sigma_n^2$ is the noise variance per dimension.
With fiber dispersion, $n(z, t) =  n(0, t) \otimes h_z(t)$ and $E\left\{n(z, t + \tau) n(z, t) \right\} = 2\sigma_n^2 \delta(\tau)$, but $E\left\{n(z_1, t + \tau) n^{*}(z_2, t) \right\}$ has a Fourier transform of $2 \sigma_n^2 e^{j \beta_2 (z_1 - z_2) \omega^2/2}$.
The temporal profile of  $\Delta u_n(t)$ can be represented by the variance of $\Delta u_n(t)$ as a function of time.
Taking into account the noise dependence, with some algebra, we find that
\begin{multline}
\sigma^2_{\Delta u_n}\!\!(t) = E\left\{ \left| \Delta u_n(t) \right|^2 \right\} = \frac{4 \gamma^2 \sigma_n^2 T_0^2A_0^4}{\pi} \\
\times \int_{-\infty}^{+\infty}
  \left| \int_{0}^L  \frac{  \exp\left( -\frac{t^2 + j \tau^2(z) \omega t + \beta_2^2 z^2 \omega^2 }  { \tau^2(z) - 2 j \beta_2 z} \!-\!\alpha z\right)}{\sqrt{\tau^2(z) - 2 j \beta_2 z}} \ud z \right|^2 \ud \omega,
\end{multline}
\noindent where $\tau(z) = \sqrt{T_0^2 + \beta_2^2 z^2/T_0^2}$ is the pulse width of \eqn{ukzt}.
Similarly, the variance profile corresponding to $j \gamma u^2_0(z, t) n^*(z, t)$ is
\begin{multline}
\sigma^2_{\Delta u_n^\prime}(t)  = \frac{\gamma^2 \sigma_n^2 T_0^2A_0^4}{\pi} \\
\times \int_{-\infty}^{+\infty}
  \left| \int_{0}^L  \frac{ \exp \left[ -\frac{(t - \beta_2 z \omega)(t - j T_0^2 \omega) }  { T_0^2 + j \beta_2 z} - \alpha z \right]}{\sqrt{T_0^4 + \beta_2^2 z^2}}  \ud z \right|^2 \ud \omega.
\end{multline}
%
%
\Fig{figsigmaut} shows the temporal profile, both the STD of $\sigma_{\Delta u_n}(t)$ and $\sigma_{\Delta u_n^\prime}(t)$ for typical fiber dispersion coefficients of $D = 17$ and $3.5$ ps/km/nm.
The initial launched pulse has an $1/e$ width of $T_0 = 5$ ps.
The fiber link is $L = 100$ km with attenuation coefficient of $\alpha = 0.2$ dB/km.
\Fig{figsigmaut} shows that the nonlinear force of $\Delta u_n(t)$ due to the beating of $|u_0(z, t)|^2$ with $n(z, t)$ is far larger than that of $\Delta u_n^\prime(t)$ due to the beating of $u^2(z, t)$ with $n^*(z, t)$.
In term of power, $\sigma_{\Delta u_n}^2(t)$ is about 1\% of $\sigma_{\Delta u_n^\prime}^2(t)$.
The noise term of  $\Delta u_n(t)$ also has more spreading over time than $\Delta u_n^\prime(t)$.
The beating of $u^2(z, t)$ with $n^*(z, t)$ can be ignored.

The temporal profile of \fig{figsigmaut} is not able to estimate the dependence between the nonlinear phase noise at $t = 0$ and, for example, $t = T$, directly.
As a trivial example for signals without chromatic dispersion and pulse distortion, the nonlinear force is proportional to $|u_0(0, t)|^2 n(0, t)$.
As white noise, the noises of $n(0, t)$ at $t = 0$ and $t = T$ are independent of each other.
In this trivial case, the profile corresponding to \fig{figsigmaut} is proportional to $|u_0(0, t)|^2$. 

If the nonlinear force of $\Delta u_n(t)$ is passing through an optical filter with an impulse response of $h_o(t)$, the filter output at the time of $m T$ is
\begin{equation}
\zeta_{0, m}  =  \int_{-\infty}^{+\infty} h_o(mT - t) \Delta u_n(t) \ud t.
\end{equation}

The SPM phase noise from $\zeta_{0, 0}$ is the noise generated by the beating of $|u_0(z, t)|^2$ with $n(z, t)$ and affect the DPSK pulse at $t = 0$.
The term of $\zeta_{0, 1}$ is IXPM phase noise from the beating of $|u_0(z, t)|^2$ with $n(z, t)$ and affect the DPSK pulse at $t = T$.
Due to IXPM, the DPSK pulse at $t = 0$ also affects by the beating of $|u_1(z, t)|^2$ (the pulse at $t = T$) with $n(z, t)$ to give the IXPM phase noise of $\zeta_{1, 0}$.
Other than the temporal location, $\zeta_{1, 0}$ is statistically the same as $\zeta_{0, -1}$.
In general, $\zeta_{k, m}$ is statistically the same as $\zeta_{0, m-k}$.
The term of $\zeta_{0, 0}$ is from SPM alone and other terms of $\zeta_{k, m}$ with $k \neq m$ is from IXPM.

Followed the model of \cite{wei03fwm, ho0504}, the differential nonlinear phase noise from both SPM and IXPM phase noise is 

\begin{equation}
\delta \phi_n = \frac{1}{A_0} \Im \left\{ \sum_m \zeta_{m, 0} \right\}
     - \frac{1}{A_1} \Im \left\{ \sum_m \zeta_{m, 1} \right\},
\end{equation}
\noindent where $\Im\{\cdot\}$ denotes the imaginary part of a complex number. To give the output of $A_0$ and $A_1$, the optical filter must have a frequency response of $H_o(\omega) = \sqrt{1 + t_o^2/T_0^2} \exp(-t_o^2 \omega^2/2)$ for the case of a Gaussian filter, where $t_o$ is the $1/e$ width of the impulse response of $h_o(t)$.

For simplicity, $A_0 = A_1$ is assumed for the same transmitted phase in consecutive symbols.
Using the property that $\Re \{ \zeta_{k, m} \}$ and $\Im \{ \zeta_{k, m} \}$ are independent and identically distributed, the variance of $\delta \phi_n$ is
\begin{equation}
\sigma^2_{\delta \phi_n} 
   = \frac{1}{A_0^2} 
    \sum_{m_1} \sum_{m_2} \left(  E \left\{\zeta_{m_1, 0} \zeta^*_{m_2, 0} \right\} 
     -   E \left\{\zeta_{m_1, 0} \zeta^*_{m_2, 1} \right\} \right) .
\label{sigmaPhiIXPM}
\end{equation}
\noindent where $\Re\{\cdot\}$ is the real part of a complex number.
 
Derived a function of $f_m(\omega)$ as
\begin{multline}
f_m(\omega)\!= 2 \gamma |A_0|^2(T_0^2\!+t_o^2)^{\frac{1}{2}} \\
\times \int_{0}^L\!\frac{ \exp \left\{ - \frac{1}{2} t_o^2 \omega^2 + \frac{\left[ (t_o^2 - j \beta_2 z)\omega  + j m T \right]^2 }  {\tau(z)^2 - 2 j \beta_2 z + 2 t_o^2}-\alpha z\right\}}{\sqrt{\tau(z)^2 - 2 j \beta_2 z + 2 t_o^2}}\! \ud z, 
\label{fmomega}
\end{multline}
\noindent  we obtain
\begin{equation}
 E \left\{\zeta_{m_1, 0} \zeta^*_{m_2, 0} \right\}
   = \frac{\sigma_n^2}{\pi} \int_{-\infty}^{+\infty} f_{m_1}(\omega) f^*_{m_2}(\omega) \ud \omega,
\end{equation}
%
%
\begin{equation}
 E \left\{\zeta_{m_1, 0} \zeta^*_{m_2, 1} \right\}
   = \frac{\sigma_n^2}{\pi} \int_{-\infty}^{+\infty} f_{m_1}(\omega) f^*_{m_2 -1 }(\omega)e^{j \omega T} \ud \omega.
\end{equation}

For an $N$-span system, the amplifier noise at the first span is the smallest and that in the last span is the largest. 
From \cite{gordon90, ho0403}, for large number of fiber spans with the identical span repeated one after another, the overall phase noise variance is $\sigma^2_{\Delta \phi_n} \approx N^3 \sigma^2_{\delta \phi_n}/3$.
The energy per pulse is $\sqrt{\pi} T_0 |A_0|^2$ with a signal-to-noise ratio (SNR) of $\rho_s = \sqrt{\pi} T_0 |A_0|^2/ (2 N \sigma_n^2)$.
The mean nonlinear phase shift is $\left<\Phi_\mathrm{NL}\right> = N \gamma L_\mathrm{eff} P_0 = N \gamma \sqrt{\pi} L_\mathrm{eff} |A_0|^2 T_0/T$ where $P_0$ is the launched power and $L_\mathrm{eff} = (1 - e^{-\alpha L})/\alpha$ is the effective fiber length. 
The variance of nonlinear phase noise of $\sigma^2_{\Delta \phi_n}$  is proportional to $\left<\Phi_\mathrm{NL}\right>^2/ \rho_s$, similar to that in \cite{gordon90, ho0403}.

\begin{figure}
\centerline{
\includegraphics[width = 0.40 \textwidth]{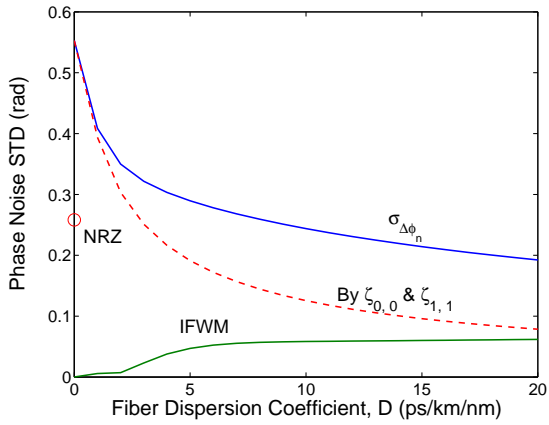}}
\caption{The phase noise STD due to nonlinear phase noise and IFWM.
The dashed-line is SPM phase noise from  $\zeta_{0, 0}$ and $\zeta_{1, 1}$ alone.
}
\label{figsigmaD}
\end{figure}

\Fig{figsigmaD} shows the phase noise STD of $\sigma_{\Delta \phi_n}$ as a function of the fiber dispersion coefficient of the fiber link.
The system has a mean nonlinear phase shift of $\left<\Phi_\mathrm{NL}\right> = 1$ rad and SNR of $\rho_s = 20$ (13 dB).
The same as \fig{figsigmaut} with $T_0 = 5$ ps, \Fig{figsigmaD} further assumes 40-Gb/s systems with $T = 25$ ps and an optical match filter of $t_o = 5$ ps.
\Fig{figsigmaD} also shows the corresponding phase STD due to IFWM calculated by the method of \cite{ho0504}.
For optical match filter, the STD from IFWM scales up by a factor of $\sqrt{3/2} = 1.22$ for the $\sqrt{3}$ times increase in the width of IFWM ghost pulse \cite{mecozzi00, mecozzi01a, ho0504}.
\Fig{figsigmaD} also includes the corresponding result for non-return-to-zero (NRZ) signal at $D = 0$ \cite{ho0403, ho0405}.

The IFWM-induced ghost pulses give a phase noise variance increase with fiber dispersion.
With large fiber dispersion and significant pulse overlap, more terms induce ghost pulses and the overall contribution from IFWM increases slowly with fiber dispersion.
From \eqn{ukzt}, fiber dispersion reduces the pulse amplitude but the increase of number of terms balances that out.
For $D > 7$ ps/km/nm, the contribution from IFWM increases slowly with the increase of fiber dispersion.

The STD from nonlinear phase noise of $\sigma_{\Delta \phi_n}$ reduces with fiber dispersion.
Even with large fiber dispersion, $\sigma_{\Delta \phi_n}$ from nonlinear phase noise is about three times larger than that from IFWM.
\Fig{figsigmaD} also shows the STD of $\Delta \phi_n$ with contribution from only SPM of $\zeta_{0, 0}$ and $\zeta_{1, 1}$.
At large dispersion, the contribution from IXPM phase noise of $\zeta_{m, k}, m \neq k$ is larger than that from SPM of $\zeta_{m, m}$.
With an interesting implication, the STD of $\sigma_{\Delta \phi_n}$ closes to that for NRZ signal at large dispersion.
The results of \cite{ho0405, ho03sta} are approximately correct for RZ pulses for system with large dispersion.

\Fig{figsigmaD} is for $N$ identical fiber spans with $\left<\Phi_\mathrm{NL}\right>  = 1$ rad.
For arbitrary link configuration, the integration of \eqn{fmomega} can be replaced by $N$ integrations for each span.
\Fig{figsigmaD} also assumes an optical match filter of $t_o = T_0$.
The function of $f_m(\omega)$ is valid for general Gaussian optical filter but other filter types are possible, may be required another layer of integration.

For lossless fiber, both \cite{green03, mckinstrie02} studied nonlinear phase noise with chromatic dispersion for continuous-wave signal \cite{green03} and without IXPM phase noise \cite{mckinstrie02}.

\section{Conclusion}

The variance of nonlinear phase noise is derived analytically for RZ-DPSK signals in highly dispersive transmission systems.
For an initial pulse width of $T_0 = 5$ ps, the phase noise STD from nonlinear phase noise is about three times larger than that from IFWM at large fiber dispersion of $D = 17$ ps/km/nm.
Nonlinear phase noise typically degrades a DPSK signal more than IFWM ghost pulses.


\end{document}